\documentclass{aa}  

\usepackage{graphicx}
\usepackage{txfonts}
\usepackage{hyperref}
\begin{document}

   \title{Adaptive Uniform Weighting: Pre-conditioning to Improve Image Fidelity}

   \author{Robert Braun
          \inst{1}
          }

   \institute{SKA Observatory, Jodrell Bank, SK11 9FT, UK
\\
              \email{robert.braun@skao.int}
             }

   \date{Received; accepted}

 
  \abstract
   {The "dirty" image made by direct Fourier inversion of visibility data is an important first step in inteferometric imaging. This is where the "deconvolution problem" is defined and the degree to which that problem is either well- or ill-conditioned has direct consequences for the ultimate image fidelity that is achieved in practise.} 
   {An under-utilised degree of freedom during Fourier imaging is the relative weights that are assigned to the visibility data. We explore the circumstances under which some adjustment of the relative weights might provide improvements to the "dirty" image, and consequently the ultimate post-deconvolution image fidelity.} 
   {We specifically explore whether typical observations with current and upcoming facilities demonstrate a significant radial trend in the acquired data density. We model those trends and use them to calculate a distinct effective local density estimate for each data point. }
   {When the resulting local density estimate is used in conjunction with a "uniform" weight correction and the desired clean beam (eg. Gaussian) tapering, it provides a significant improvement in the image quality over that provided by the current pixel-based density estimate. This distinction disappears in those cases where the acquired visibility sampling is essentially complete. }
   {In many cases, particularly spectral-line observations and those with only limited sidereal tracking, this adaptive approach improves the beam quality by a factor of 2 to 10, as measured by the RMS residual relative to the best-fitting clean beam, providing an improvement in final image fidelity that is similar in magnitude. An appealing aspect of this approach is that there are no "knobs" for the user to adjust. Once the field size and pixel size are specified (guided by astrophysical aims) the method is fully adaptive to the data that has been acquired and produces the "cleanest possible" dirty beam for these circumstances.}


   \maketitle

\section{Introduction}
\label{sec:intro}
Interferometric imaging in radio astronomy \citet{thompson} has been utilised now for more than 60 years as a means to achieve higher angular resolution than is possible with monolithic single dish radio telescopes. Although the basic principles are well-understood, there are still areas where the practical implementation of those principles might benefit from further analysis. One in particular that we will consider here is the choice of the data weighting that is applied to the measured visibilities. This is a topic that, despite its importance in determining image properties, such as thermal noise, source confusion and image fidelity has received very little study. One reason for this lack of interest in optimising the data weighting might be that it is considered only of marginal relevance since it applies to the "dirty" interferometric image, which must inevitably be deconvolved to achieve a "clean" outcome prior to any quantitative analysis. In fact, it is actually of fundamental importance to achieve the "cleanest possible dirty image" by linear means during the imaging process itself. In this way, the image flux scale is best preserved by matching beams of any modelled emission with the inevitably unmodelled image residuals. And perhaps of even greater importance is the issue of image fidelity. As we will demonstrate below, the "cleanest possible dirty beam" produced by only linear means, provides quite a good indication of the intrinsic "image fidelity", the normalised difference between intrinsic and reconstructed image structures after deconvolution. Once non-linear methods are introduced into the imaging problem they bring with them inherent biases and assumptions about the underlying sky brightness, that have been designed to make plausible looking images, but tend not to be reproduced once higher resolution and sensitivity observations are in hand. 

\section{The Imaging Problem}
\label{sec:problem}
The challenge one faces in interferometric imaging is estimation of a robust, high-fidelity gridded representation of the sky brightness, using incomplete and noisy measurements of its Fourier Transform, that is suitable for quantitative analysis. Since the field of view is often large relative to the angular resolution, it can encompass $>10^6$ distinct sources above the noise floor, all of apriori unknown detailed morphology. This makes it vital that the resulting "clean" image has a well-defined, position-invariant PSF that can be unambiguously represented on both the imaging grid, and (ideally) its Fourier Transform, particularly for sources that will inevitably not be centered on image pixels. For a given pixel size, $\delta$, an obvious choice for the "clean" PSF is a Gaussian with FWHM $\gtrsim 2.5\delta$. Other functional forms might also be considered for the "clean" PSF if they satisfy these requirements and brought additional benefits.

\section{Data Weighting}
\label{sec:weights}
When considering the impact of visibility data weighting on the PSF we need only consider the small angle approximation, where the line-of-sight $w$ coordinate of projected baseline length is ignored,  since the three dimensional treatment is only required to make the bore-sight PSF invariant across the field, as we will explicitly verify below. As outlined in \citet{thompson}, the measured visibility data, $V_{meas}(u,v)$ can be related to the intrinsic visibility data, $V(u,v)$, as, 
\begin{equation}
\label{eqn:V_meas}
    V_{meas}(u,v) = W(u,v) \, w(u,v) \, V(u,v),
\end{equation}
where $W(u,v)$ represents the transfer function that expresses the visibility sampling achieved by the instantaneous interferometric array configuration in conjunction with any sidereal tracking. Data weighting refers to the term, $w(u,v)$, in the equation above. The measured "dirty" image is given by the Fourier transform of that equation,
\begin{equation}
    I_{meas}(l,m) = I(l,m) * b_0(l,m),
\end{equation}
where the asterisk, $*$, represents a two dimensional convolution and $b_0(l,m)$, is the "dirty" beam, which is the Fourier transform of $W(u,v) \, w(u,v)$. Historically, there have only been two basic data weighting methods used, termed "natural" and "uniform". Natural weighting makes use of the inverse data variance for $w(u,v)$ and results in images with the lowest possible receiver noise fluctuation level, although often at the expense of very undesirable dirty beam properties for typical array configurations. In fact, in many circumstances, the natural PSF can have a both extremely broad side-lobes that scatter even compact source power out to large angular distances as well as a central cusp that is under-sampled relative to the imaging grid, which creates an ill-conditioned imaging problem. Uniform weighting makes use of the inverse data density, 
\begin{equation} 
    w_u(u,v) \propto 1/\rho(u,v), 
\end{equation}
most often in conjunction with a tapering function $w_t(u,v)$, that represents the form of a clean "restoring" beam, so that,
\begin{equation}
    w(u,v) = w_u(u,v) \, w_t(u,v),
\end{equation}
where the tapering is normally chosen to be a Gaussian, due to its many desirable properties, as we had noted in the previous section. The most useful Gaussian tapers are those that will satisfy the Nyquist-Shannon sampling theorem (corresponding to an image plane PSF with FWHM $\gtrsim 2.5\delta$) and decline to small values at the largest radii where data have been acquired, so that no extrapolation to larger radii (by non-linear deconvolution methods) is made necessary. 

A third weighting option was developed by \citet{briggs} in which a tunable "robustness" parameter allows the user to choose some linear combination of natural and uniform weighting factors to yield intermediate beam and noise properties between the two. This approach provides very little control of the resulting dirty PSF shape and could also lead to a PSF that is not fully sampled on the imaging grid.

A final variant that has been implemented in some imaging software is termed "super-uniform" weighting. In this case, the inverse data density factor is evaluated on a coarser grid than that used for imaging, typically an integer multiple of the original pixel size. Although this can sometimes be beneficial, there are also some potential pitfalls with this approach, as we will discuss below.

\begin{figure*}
   \centering
    \includegraphics[bb=91 232 592 690,clip, width=17cm]{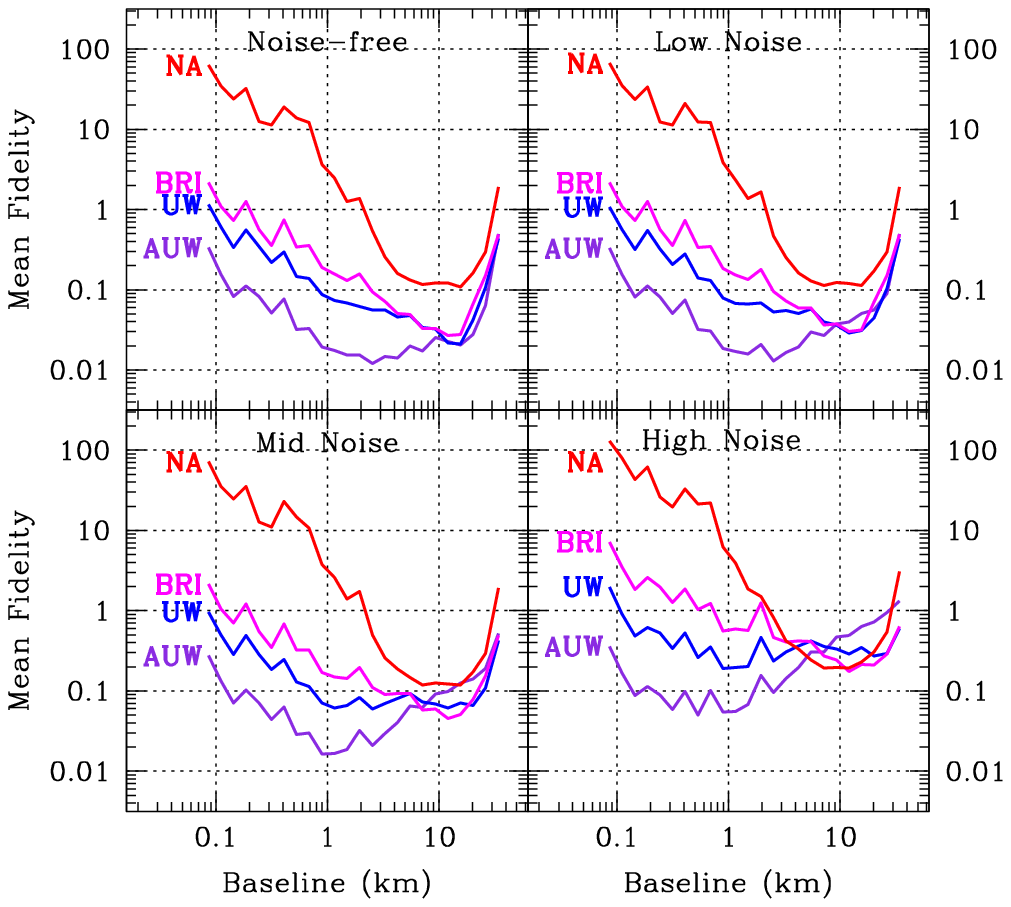}
      \caption{Comparison of restored image fidelity with standard Uniform Weights (UW), Briggs, (BRI), Natural (NA) and Adaptive Uniform Weights (AUW) for a noise-less (top-left), low-noise (top-right), nominal-noise (bottom-left) and high-noise (bottom-right) monochromatic 4-hour tracking observation with SKA-Low. Image fidelity has been averaged in logarithmically spaced annuli. }
        \label{fig:Fidelity}
   \end{figure*}

\begin{figure*}
   \centering
    \includegraphics[width=17cm]{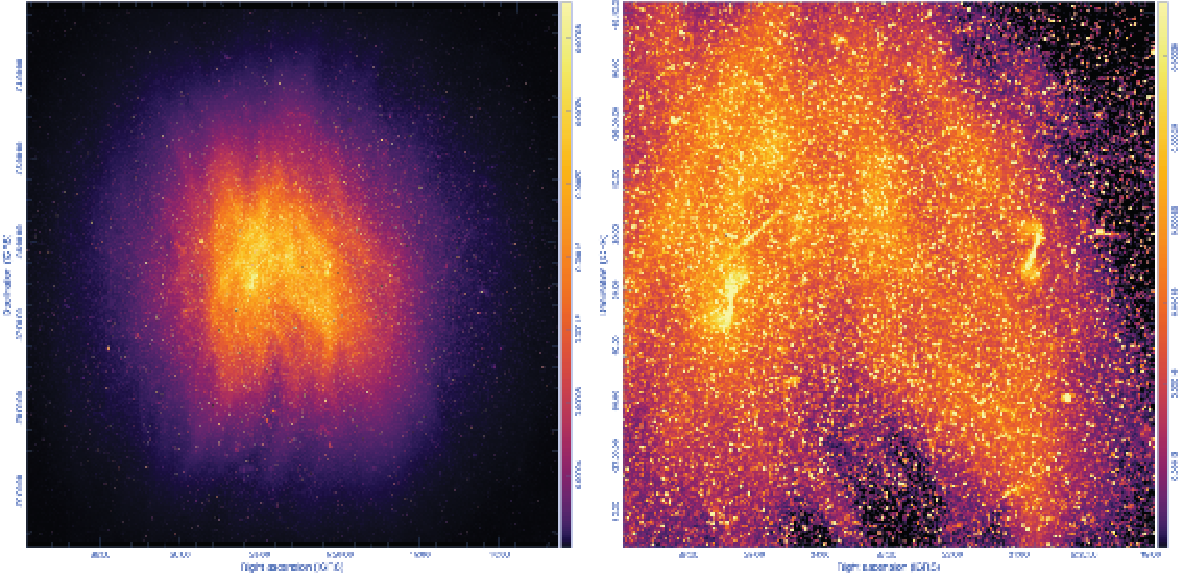}
      \caption{Sky model used for image fidelity simulations. The full $6.4\times 6.4$ degree field tapered with the primary beam model is shown on the left and the central portion on the right. The peak brightness is 1.5 Jy/Beam, the diffuse emission plateau has brightness of 0.2 mJy/Beam (about 25 K), the faintest discrete sources are < 1 $\mu$Jy/Beam.}
        \label{fig:SkyModel}
   \end{figure*}

We provide some motivation for exploring alternative data weighting approaches by contrasting the image fidelity of "restored" deconvolved images resulting from "dirty" images made with the Natural, Briggs (using \textit{miriad} robust = 0), Uniform, as well as the Adaptive weighting strategy defined here for a monochromatic, four-hour tracking observation with SKA-Low in Fig.\ref{fig:Fidelity}. We have explored a wide range of signal-to-noise scenarios, including noise-free, low-noise (nominal/3), nominal noise and high-noise (nominal$\times$5). Image fidelity is defined here as,
\begin{equation}
    F = \frac {|\mathcal{F}(S - R)|} {|\mathcal{F}(S)|},
\end{equation}
the magnitude of the Fourier Transform, $\mathcal{F}$, of the difference between the ideal sky model, $S$, and the restored sky model, $R$, normalised by the ideal sky model magnitude. The image fidelity has been averaged in logarithmically spaced annuli to demonstrate its variation with scale. The sky model, imaging and deconvolution simulations are described in Sect. \ref{sec:fidelitysims} and the sky model itself is shown in Fig. \ref{fig:SkyModel}. What is apparent from Fig.\ref{fig:Fidelity} is that the resulting image fidelity is closely tied to the quality of the "dirty" image, ie. the degree to which the dirty PSF is already consistent with the ultimately desired restoring function for the image. The only circumstance to which this does not apply, is when the signal-to-noise ratio is so low that source "detection", rather than quantitative characterisation is being attempted. As expected, Natural weighting is unsurpassed for source detection, but is poorly suited to complex source characterisation.

\section{Data Density}
\label{sec:density}
A vital quantity that underlies the uniform weighting strategy is the data density, $\rho(u,v)$. While conceptually simple, this quantity can only be specified analytically in a few special cases, such as a linear, regularly sampled, east-west array undertaking a full 12 hour tracking observation at high declination \citep{thompson} for which $\rho(u,v) \propto 1/(u^2+v^2)^{1/2}$. For a more typical array configuration and track duration the estimation of local data density becomes much more problematic. 

The most common implementation of data density estimation now in use makes use of the data grid that is defined for Fourier transform imaging. Prior to the actual data gridding step, which involves convolution to a grid with a suitable gridding kernel while potentially taking explicit account of the $w$ coordinate, the weights that will be applied to the data are first determined in only the $(u,v)$ plane. 
A simple sum of data points that fall (in the nearest pixel sense) within each $(u,v)$ cell provides the local density estimate. In the event that the majority of $(u,v)$ cells contain at least one data point, then this is an excellent estimate. However, this condition is rarely met in practise. If one considers imaging with a relatively fine pixel sampling and only short observation duration, then one can often approach the regime where each $(u,v)$ cell has only either a single data value or none at all and the overall fraction of filled cells is rather low. In this case, there is little distinction between natural and uniform weighting, and all of the resulting dirty beams are equally bad.

\section{Adaptive Uniform Weighting}
\label{sec:AUW}
We have explored an alternative strategy for local density estimation to utilise in a uniform weighting correction. In doing so, we make a clear distinction between a factor needed to account for multiple measurements within the same cell, $w_m(u,v)$, and the factor needed to account for the local density of occupied cells, $w_o(u,v)$. The uniform weight correction factor is the product of the two, 
\begin{equation}
w_u(u,v) = w_m(u,v) \, w_o(u,v).
\end{equation}
We start by forming the gridded distributions of data multiplicity, $M(u,v)$ and data occupancy, $O(u,v)$, while also determining the minimum and maximum radii, $r_{min}$ and $r_{max}$ (in units of pixels) where data occur. It is important to note that while the multiplicity distribution can have values of either 0, or any positive value, the occupancy distribution has only values of either 0 or 1. Next we consider whether there might be any systematic variation of the average occupation density as function of radius in the visibility plane, $o(r)$. There are at least two reasons why such a trend might be present. Firstly, many array configurations tend to be centrally concentrated to provide high surface brightness sensitivity, so that there is often a radial decline in the number of sampled $(u,v)$ points. Secondly, as noted in the previous section, the presence of any sidereal tracking also gives rise to a natural radial decline in the data occupancy. 

We have considered the distributions of $o(r)$ for many array configurations of current facilities: ALMA \citep{cortes}, LOFAR \citep{haarlem}, MeerKAT \citep{jonas}, the VLA \citep{napier} as well as the the upcoming SKA-LOW and SKA-MID \citep{braun}. What we find is a wide variety of behaviour, which can be summarised as follows. There is a peak occupancy, $o_p$, at some value of $r = r_p$ and there is a possible roll-off of average occupancy both to larger and in some cases smaller radii. Some examples of average occupancy are provided below in Sect. \ref{sec:implementation}.

The strategy we explore is that for radii with average occupancy larger than some cut-off value $o(r) \ge o_{t}$ we simply adopt the standard Uniform weighting approach, where local occupancy is equal to the gridded occupancy,
\begin{equation}
    o(u,v) = O(u,v),
\end{equation}
while for radii where $o(r) < o_{t}$ we undertake an alternate local occupancy density estimation. This is accomplished by using the average occupancy at each radius to define a smoothing kernel diameter,
\begin{equation}
\label{eqn:d_of_r}
    d(r) = T / o(r)^{1/2},
\end{equation}
that is a factor of $T$ larger than what yields an average of one occupied cell per kernel. Each occupied cell of $O(u,v)$ is convolved with the smoothing kernel, $k(u,v)$, chosen for example to be a uniform disk with diameter, $d(r)$, and normalised to an integral of unity,
\begin{equation}
o(u,v) = O(u,v) * k(u,v).
\end{equation}
This is the local occupation density estimator that provides $w_o(u,v) = 1/o(u,v)$ to be used in conjunction with the multiplicity factor $w_m(u,v) = 1/M(u,v)$ in (5). We have explored alternate smoothing kernel definitions, but found no benefit in using more complex kernel shapes.

The reader will have noted that the variable, $T$, was introduced in (7), to ensure that the smoothing kernel should span multiple occupied cells. This is necessary to accommodate potential variability in the occupation density with azimuthal angle at each radius. It was found that a value $T \sim 3$ seems to perform best in terms of the resulting dirty beam quality.

   \begin{figure*}
   \centering
    \includegraphics[bb=20 250 581 693,clip, width=15cm]{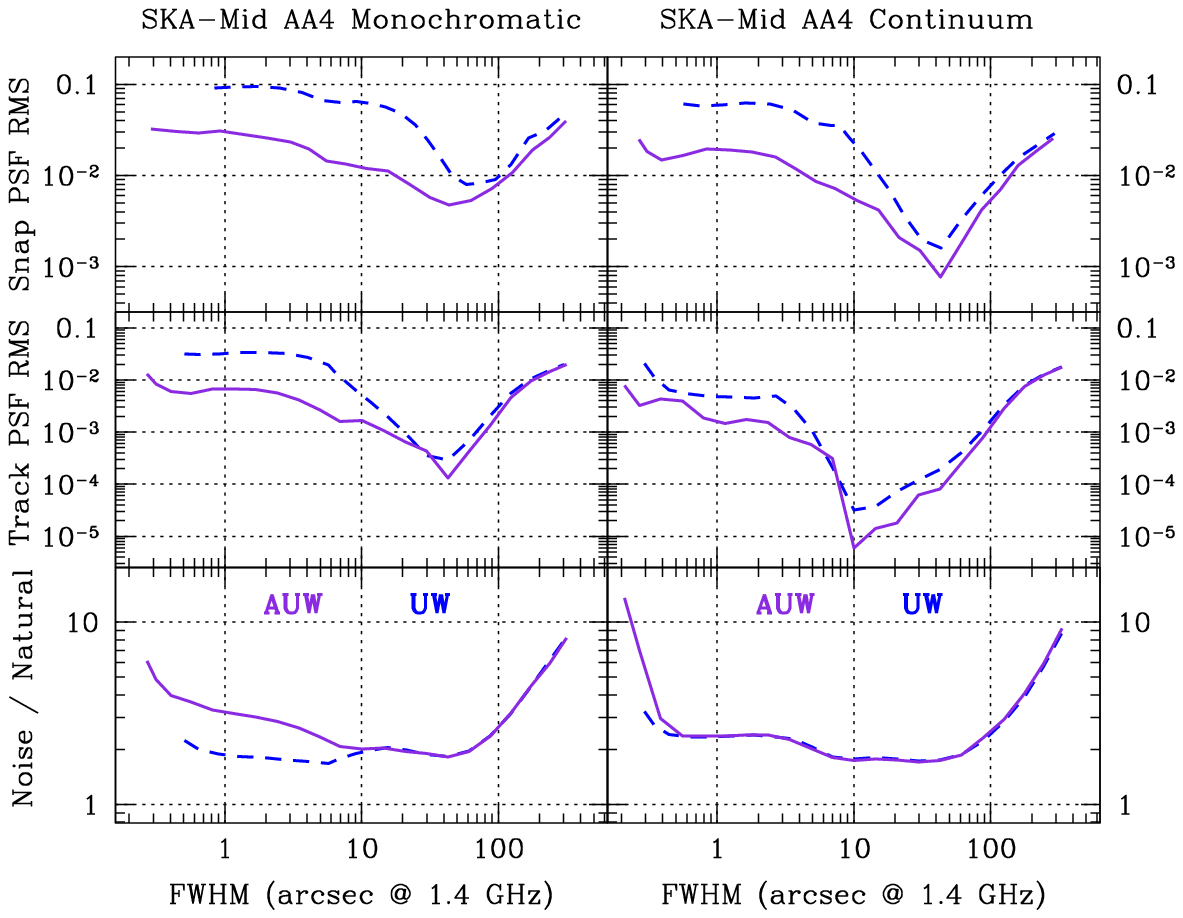}
      \caption{Comparison of imaging performance with standard Uniform Weights (UW, dashed lines) and Adaptive Uniform Weights (AUW, solid lines) for SKA-Mid. The monochromatic case is on the left and multi-frequency synthesis with 40\% fractional bandwidth on the right. Bottom panels show the image RMS noise relative to Natural weighting. The other panels show the RMS residual after a Gaussian fit to the PSF for a snap-shot (top) and full track (centre) observation.}
        \label{fig:SKA-Mid}
   \end{figure*}

\section{Relative PSF Performance}
\label{sec:perform}
We have implemented the "adaptive uniform weighting" (AUW) method, using the detailed implementation outlined in Sect. \ref{sec:implementation}, within the $miriad$ package \citep{sault} imaging task $invert$ to explore its utility and contrast its performance with the standard pixel-based density estimator of "uniform weighting" (UW). We have made use of the SKA-Mid and SKA-Low Design Baseline array configurations, as well as those of MeerKAT, the VLA-A configuration, LOFAR and ALMA (in its Cycle 10-8 configuration). We explore both transit snap-shot and sidereal track (for Dec. = $-/+30^\circ$ and $-4^h < HA < +4^h$ for the mid-frequency dish arrays and $-2^h < HA < +2^h$ tracks for aperture arrays and ALMA) imaging for both monochromatic and multi-frequency synthesis  (MFS with $\Delta\nu/\nu = 0.4$) continuum observations. For the sidereal tracking simulations we adopt a sampling interval of 1 minute, rather than the shorter time interval that would represent continuous data acquisition at a rate which minimises time smearing. This is done to keep the simulation times and data quantities more tractable.

For each simulated observation we explore the full range of angular scales that the array configuration provides, by defining a logarithmic sequence of target Gaussian FWHM beams, $\theta_t$, that define the Gaussian tapering function $w_t(u,v)$ at an arbitrary reference frequency accessible to that array. The pixel size is chosen to be $\theta_t/4$ and the image size is chosen to ideally span a diameter of $1.5 \times$ the first null of the primary beam pattern although it is also constrained to be in the range $1024 \le N_{pix} \le 16384$ pixels.

We illustrate the relative PSF properties of Adaptive Uniform Weighting (AUW) versus standard Uniform Weighting (UW) in Figure \ref{fig:SKA-Mid} for SKA-Mid. For the standard UW results we have undertaken the same imaging simulations with both $miriad$ \citep{sault} as well as $wsclean$ \citep{offringa}. While both produce similar results, the $wsclean$ images demonstrate a more robust continuity with imaging parameters and are used for all of the comparisons below. 

The SKA-Mid array configuration provides good sensitivity, as measured by the ratio of RMS image noise relative to the Natural image noise, to an extremely wide range of angular scales (about 1000:1) as can be seen in the bottom panels. The quality of the dirty PSF, $Q$, is measured by the RMS residual within the central $16 \times 16$ beam FWHM after subtracting the best-fitting Gaussian, and is shown in the other panels of the Figure. Not surprisingly, better PSF quality is achieved in sidereal tracking observations relative to snap-shots, as well as MFS observations relative to monochromatic ones. In those cases where the visibility sampling is already essentially complete, there is no distinction between the UW and AUW PSF. The distinction becomes apparent when higher angular resolution is targeted, in this case below about 10 arcsec (at a reference frequency of 1.4 GHz) and is largest for the monochromatic and snap-shot coverage cases, where there can be up to an order of magnitude lower PSF RMS. The more Gaussian PSF provided by AUW imaging is also accompanied by more effective tracking of the target Gaussian FWHM in each simulation, extending the highest angular resolution achievable by a factor of about 1.5, up to the diffraction limit of the array. The improved and more compact PSF, particularly for the monochromatic case, does come at the expense of a somewhat enhanced RMS image noise, as seen in the lower left panel of the Figure. 

The same AUW versus UW comparison is shown in Figures \ref{fig:LOFAR}  (LOFAR), \ref{fig:SKA-Low}  (SKA-Low),  \ref{fig:MeerKAT} (MeerKAT),\ref{fig:VLA-A} (VLA) and \ref{fig:ALMA} (ALMA). Very similar trends to those already noted for SKA-Mid can be seen in these cases. The most significant improvements in PSF quality are obtained for monochromatic observations as well as for less complete sidereal tracking.

\section{Discussion}
\label{sec:discuss}
\subsection{Benefits}
As noted in Sect. \ref{sec:weights} and illustrated in Fig.\ref{fig:Fidelity}, the primary motivation for improving the quality of the dirty PSF comes from enhancing the image fidelity that will ultimately be achieved following any required deconvolution. What is apparent from Fig.\ref{fig:Fidelity} is a significant improvement in final image fidelity over the complete range of angular scales provided by the array configuration that scales with the quality of the dirty PSF. This generalisation only breaks down in the noise-dominated regime, where the higher RMS noise that accompanies an improved dirty PSF begins to limit source detectability relative to the more nearly Natural visibility weightings. To put this series of simulations into context (a monochromatic $4^h$ tracking observation with SKA-Low) it is useful to consider the left-hand-central panel of Fig.\ref{fig:SKA-Low}. The PSF RMS for the AUW and UW cases is about $Q=$ 0.01 and 0.03 respectively for a target FWHM near 10 arcsec. This is comparable to the best noise-free image fidelity, $F$, that is achieved at intermediate angular scales for these two weighting methods in our simulations. As the image signal-to-noise level decreases, image fidelity deteriorates from this starting point. What this sequence of simulations demonstrates is that the dirty PSF RMS, $Q$ provides a plausible best-case (high signal-to-noise, negligible calibration error) estimate of the final deconvolved image fidelity, $F$. It is clearly very desirable to achieve the cleanest possible dirty image. 

\subsection{Optimisation}
The specific implementation of the AUW strategy described in Sect. \ref{sec:implementation} has been shown to be moderately effective in Figures \ref{fig:SKA-Mid}, \ref{fig:LOFAR}, \ref{fig:SKA-Low}, \ref{fig:MeerKAT}, \ref{fig:VLA-A} and \ref{fig:ALMA}. However, there is undoubtedly scope for further optimisation of this approach. We have employed a discrete convolution to obtain a local estimate of occupation density at radii where the azimuthally averaged density declines below unity. While this has an acceptable computational cost for moderate visibility numbers, it may be more efficiently undertaken with a sequence of FFT-based convolutions using Gaussian tapers that sample the required range of $(u,v)$ kernel sizes. 

Another indication that further optimisation is needed comes from the occasionally discontinuous behaviour of the resulting PSF properties with the gridding parameters. A noteworthy case is the full-track MeerKAT MFS case shown in Figure \ref{fig:MeerKAT}, where the PSF RMS demonstrates some oscillatory behaviour when targeting FWHM in the range 4 to 20 arcsec. 

\subsection{Relation to Super-Uniform Weighting}
It is worth noting that the occurrence of an average data occupancy of less than unity is exactly what motivates the use of "super-uniform" weighting. However, while the adaptive approach attempts to track the spatial variation of occupancy density, the super-uniform method employs a single, typically square "smoothing" kernel for the entire visibility grid. In this context it is also worth drawing attention to the fact that those implementations of super-uniform weighting that simply employ a coarser "super-grid" that consolidates $n\times n$ original $(u,v)$ pixels into one, are prone to generate aliased secondary peaks in the PSF at multiples of $1/n$ of the field size in both dimensions. A robust implementation of super-uniform weighting is one in which the gridded weights are instead subjected to convolution with an appropriate kernel, for example a Gaussian with FWHM of several $(u,v)$ cells. This robust method of super-uniform weighting has been implemented within the $miriad$ package as the "radfft" option within the $invert$ imaging task.

\subsection{Application to the full 3D Imaging Case}
As noted at the outset, the strategy outlined here for visibility weight determination has deliberately considered the case where the $w$ coordinate of the visibility data can be ignored. For wide-field imaging at relatively low radio frequencies this is rarely the case. However, once the visibility weights have been determined for an optimised bore-sight PSF, as outlined in this manuscript, they can used in conjunction with any desired 3D imaging technique, such as $w$-projection \citep{cornwell} or $w$-stacking \citet{offringa} to achieve the necessary degree of position invariance of the PSF across the field. We have tested this assertion explicitly, by implementing the algorithm described in Sect. \ref{sec:implementation} as a stand-alone python script that provides pre-processing of a visibility database and produces a MeasurementSet (MS) that can be imaged and deconvolved with other applications. The visibility data weights are optimised for a specific desired image (in terms of field size, pixel sampling and possible multi-frequency synthesis). The MS was then imaged with \textit{wsclean} by specifying Natural weighting together with a Gaussian tapering appropriate for the image pixel size and this demonstrated accurate field invariance (as specified by the \textit{-wgridder-accuracy} parameter) of the bore-sight PSF including its excellent AUW attributes.

\begin{acknowledgements}
The assistance and support of Mark Wieringa during the implementation of this strategy into the $miriad$ package $invert$ task as the weighting option "radial" is greatly appreciated, as is his implementation of an FFT-based "super-uniform" weighting option "radfft" within $invert$.    
\end{acknowledgements}

\newpage
\begin{appendix}
\section{Image Fidelity Simulations}
\label{sec:fidelitysims}
Here we describe the imaging and deconvolution simulations undertaken to quantify image fidelity.

We constructed a realistic sky model, $S$, for relatively low radio frequencies (180 MHz as an example) and Southern declinations at relatively high Galactic latitude by beginning with the GLEAM survey \citep{GLEAM} complete to 100 mJy and supplementing this with a T-RECs simulation \citep{bonaldi} with a statistically accurate representation of extragalactic source numbers and morphologies between 100 mJy and 1 $\mu$Jy (about 1.5 $\times 10^7$ in number). A plausible diffuse Galactic foreground for the field was drawn from an MHD simulation of the ISM \citep{bracco} and this was arbitrarily assigned a peak brightness of about 25 K, consistent with expectations for a high latitude field. The complete model spanning 6.4 degree, was gridded with a 12.5 arcsec FWHM circular Gaussian on a 4 arcsec grid and spatially tapered with a model of the primary beam, including its first side-lobe, prior to being zero-padded to double the pixel number in both dimensions (11520 $\times$ 11520). The resulting sky model is shown in Fig. \ref{fig:SkyModel}

Dirty beams, $b$, and noise images, $N$, were generated from a simulated monochromatic 180 MHz observation with \textit{miriad uvgen} spanning hour angles of $-2^h$ to $+2^h$, sampled at intervals of $15^m$ for the SKA-Low configuration of 512 stations with $B_{max} \sim 74$km. \textit{miriad invert} was used to produce the dirty beams with Natural, Briggs (robust = 0), Uniform and Adaptive Uniform (options = radial) weighting sampled by (11520 $\times$ 11520) pixels of 4 arcsec. The Briggs, Uniform and Adaptive Uniform beams were generated with a Gaussian taper of 12.5 arcsec FWHM, while the Natural beam was left untapered.

A deconvolved sky model representation, $S^\prime$, was generated from the (11520 $\times$ 11520) Gaussian gridded model by a linear deconvolution of a unit height 12.5 arcsec FWHM Gaussian, $G$, truncated at $1\times10^{-6}$,
\begin{equation}
    S^\prime = \mathcal{F}^{-1}\bigg(\frac{\mathcal{F}(S)}{\mathcal{F}(G)}\bigg).
\end{equation}
Convolved dirty images, $D$, were constructed as,
\begin{equation}
    D = \mathcal{F}^{-1}\big(\mathcal{F}(S^\prime)\ \mathcal{F}(b)\big) + N.
\end{equation}

The dirty images and beams were used to undertake a deconvolution of the field with \textit{miriad clean}. All deconvolved models were restored with a 12.5 arcsec FWHM Gaussian that was added to the relevant residuals. The deconvolution was undertaken both in the absence of noise, and with low-, nominal- and high-noise values; Natural RMS noise  of $\sigma_{NA} = 0, 41, 124, 620 \  \mu$Jy/Beam. The corresponding RMS noise values for the Briggs, UW and AUW cases are higher than the Natural values by factors of 1.35, 1.67 and 2.41 respectively in all cases. For most cases (noise-free, low- and nominal-noise), a target brightness limit of $1\times10^{-4}$ Jy/beam or iteration number limit of $10^6$ was specified. For the high-noise case a target brightness limit of $5\times10^{-4}$ Jy/beam or iteration number limit of $2\times10^5$ was specified. The Briggs, Uniform and Adaptively weighted images were all able to reach the requested brightness limit, although this required fewer iterations with UW and fewer still for AUW. The Naturally weighted deconvolutions made little net progress and terminated after the full iteration number without reaching the target brightness limits in any of the cases considered. 

\section{Implementation Strategy}
\label{sec:implementation}
Here we outline the specific implementation of the Adaptive Uniform Weighting strategy within the $miriad$ \citep{sault} package imaging task $invert$. 

As noted in Sect. \ref{sec:weights}, the gridded distributions of data multiplicity $M(u,v)$ and occupancy $O(u,v)$ are first accumulated on the basis of data assignment to only the nearest pixel in the grid. At the same time the minimum and maximum radii, $r_{min}$ and $r_{max}$, (in units of pixels) where data occur are also determined, while imposing the constraint that $r_{max} \le N_{pix}/2$. This is done to preserve circular symmetry in the resulting beam shape.

We then define a set of radial annuli with a logarithmic radial step. We arbitrarily choose a number of bins $n_r = (r_{max}-r_{min})^{1/2}$, constrained to be $n_r \le 64$ that define a set of outer radii, $r_i$ given by,
\begin{equation}
    r_i = antilog_{10}[log_{10}(r_{min}) + i \times log_{10}(r_0)], \, \, \, i = 1, ... n_r,
\end{equation}
where, 
\begin{equation}
    r_0 = antilog_{10}[log_{10}(r_{max}) - log_{10}(r_{min}))/n_r].
\end{equation}
We then determine the average cell occupancy, $0 \leq o_i \leq 1$, within each annulus between, $r_i$ and $r_{i-1}$ with indicative radius $r_i' = (r_i + r_{i-1})/2$. 

We have adopted the following functional form to describe the variation of average occupancy with radius:
\begin{eqnarray}
    o(r) & = & o_p/[1+(r/r_2)^{\alpha_2}] \quad r > r_p \\
    & = & o_p/[1+(r_1/r)^{\alpha_1}]  \quad r < r_p,   
\end{eqnarray}
where the possible roll-off from the peak to both larger and smaller radii is described by a power-law. The peak value $o_p$ is first defined as the 95$^{th}$ percentile of the $o_i$ distribution, while its location in the sequence of radii is identified as $i_p$.

Trial power-law indices $\alpha_1$ and $\alpha_2$ with discrete values $\alpha_j = 0.1,0.2,\ldots 10.0$ were used to determine the corresponding values of $r_1$ and $r_2$. For each trial power law index, the weighted average value of the corresponding reference radius, $r_{1}(j)$ or $r_{2}(j)$, was determined over the relevant radial bins (either those below or beyond the peak),
\begin{eqnarray}
r_1(j) & = & \sum_i w_i {r_i \over (o_p/o_i - 1)^{\alpha_j}}  \Bigg/ \sum_i w_i \quad i < i_p \\
r_2(j) & = & \sum_i w_i  r_i (o_p/o_i - 1)^{\alpha_j}  \Big/ \sum_i w_i \quad i > i_p    
\end{eqnarray}
where the weights, $w_i$ are given by the total number of cells in each annulus. Only those radial bins for which $o_i/o_p < 0.98$ were used to constrain the fits. The best-fitting combinations of $(\alpha_1,r_1)$ and $(\alpha_2,r_2)$ are then identified by determining which provide the minimum RMS residual relative to all of the measured $o_i$ for $i < i_p$ and $i > i_p$ respectively.

\begin{figure*}
   \centering
    \includegraphics[bb=20 250 581 693,clip, width=17cm]{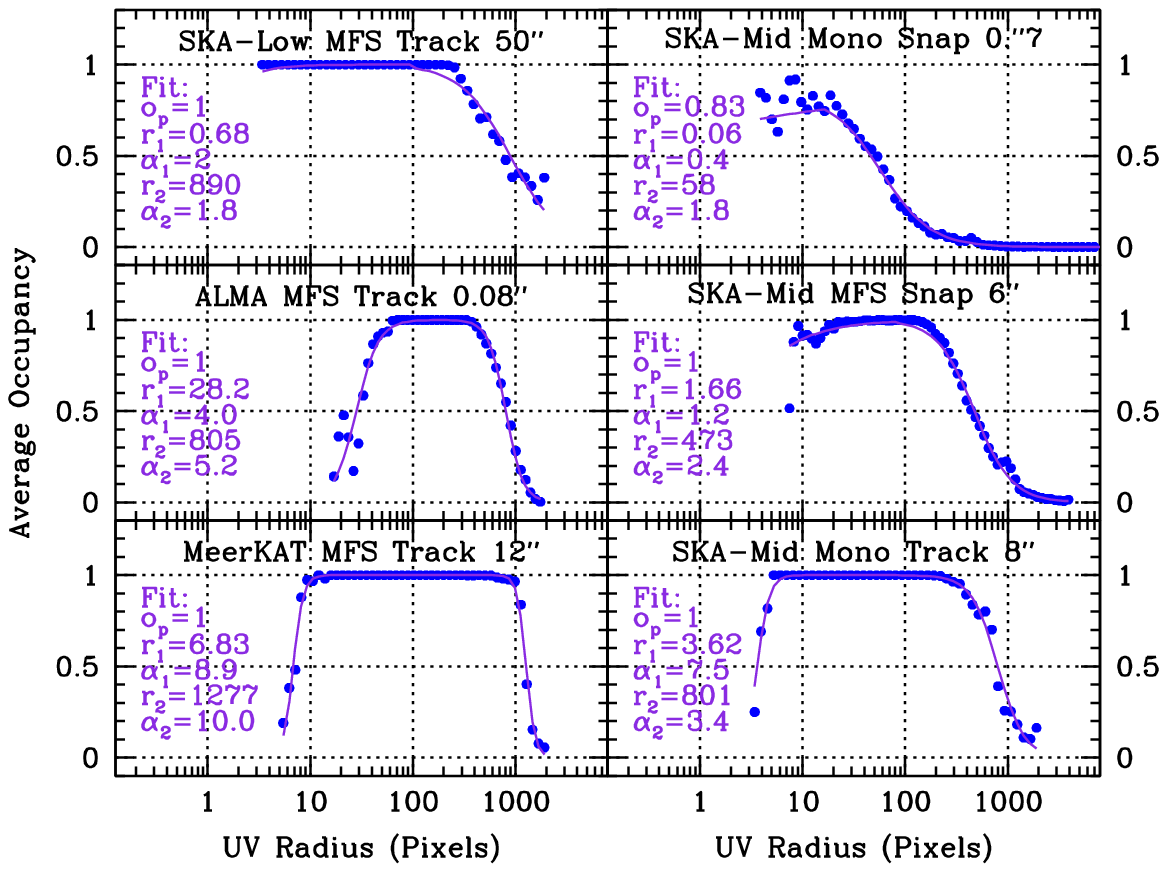}
      \caption{Examples of average occupancy versus $(u,v)$ radius (filled circles) for different facilities, observing strategies and target beam FWHM. The fit to $o(r)$ is overlaid as the solid line and parameter values are listed. }
        \label{fig:o_of_r}
   \end{figure*}

If no valid fit parameters are found due to the average occupancy not declining below $0.98 o_p$, then default values are assigned to produce a flat $o(r)$ dependence below and/or above the peak. We provide a few illustrative examples of actual $o(r)$ distributions together with the model fits in Fig. \ref{fig:o_of_r}. As is apparent from the figure there is great diversity in the distributions which varies with the spectral and time sampling in conjunction with the specific antenna layout of each facility. 

With this numerical model of the average occupation density in hand we return to the gridded occupancy distribution, $O(u,v)$, calculate the corresponding smoothing kernel diameter as given in eqn. \ref{eqn:d_of_r}, which we arbitrarily constrain to be no larger than, $d(r) \le 129$ pixels, and continue as described in the final paragraphs of Sect. \ref{sec:AUW}.

\section{Other Facilities}
\label{sec:facilities}
Here we demonstrate the improvement in dirty PSF properties when using the AUW strategy for a variety of other upcoming and existing synthesis imaging arrays.

   \begin{figure*}
   \centering
    \includegraphics[bb=20 250 581 693,clip, width=14.5cm]{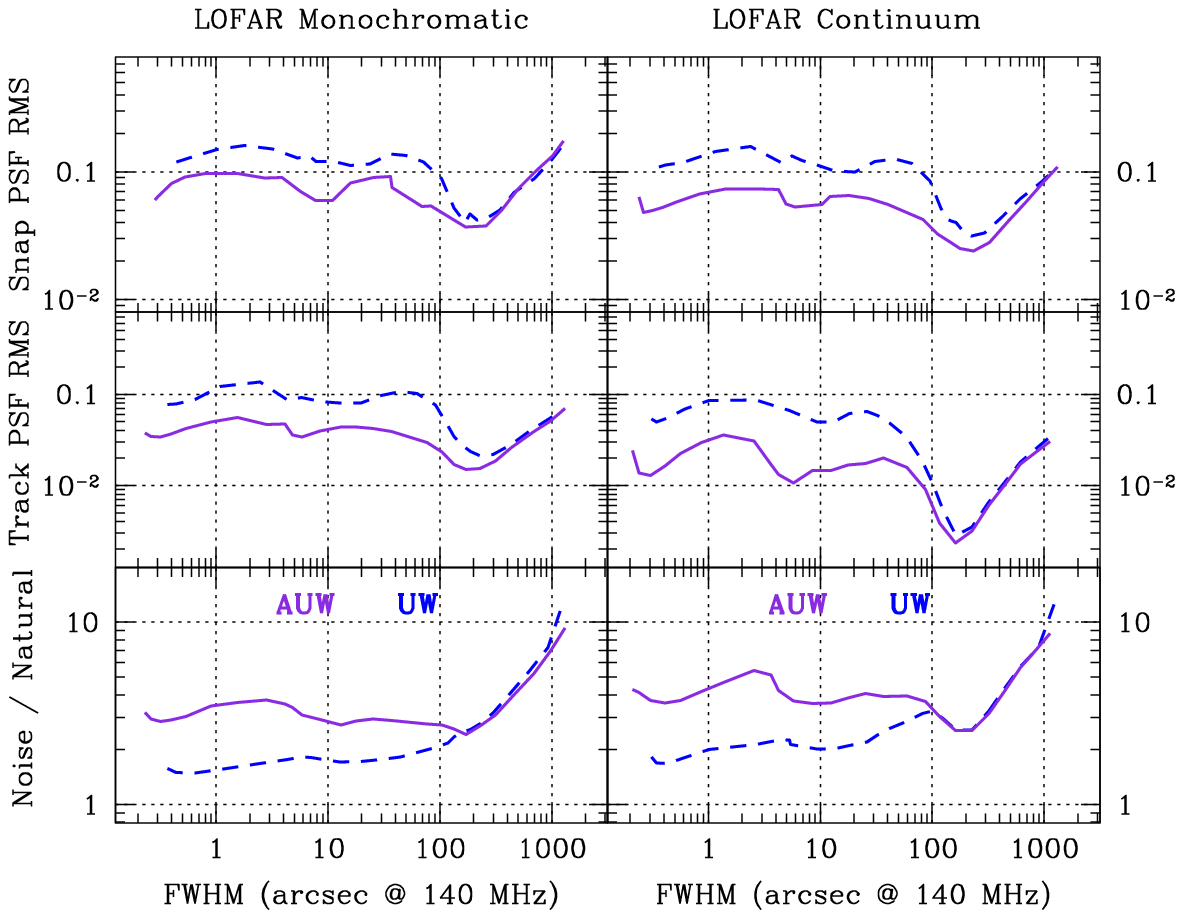}
      \caption{As in Figure \ref{fig:SKA-Mid}  for the International LOFAR telescope.}
        \label{fig:LOFAR}
   \end{figure*}

   \begin{figure*}
   \centering
    \includegraphics[bb=20 250 581 693,clip, width=14.5cm]{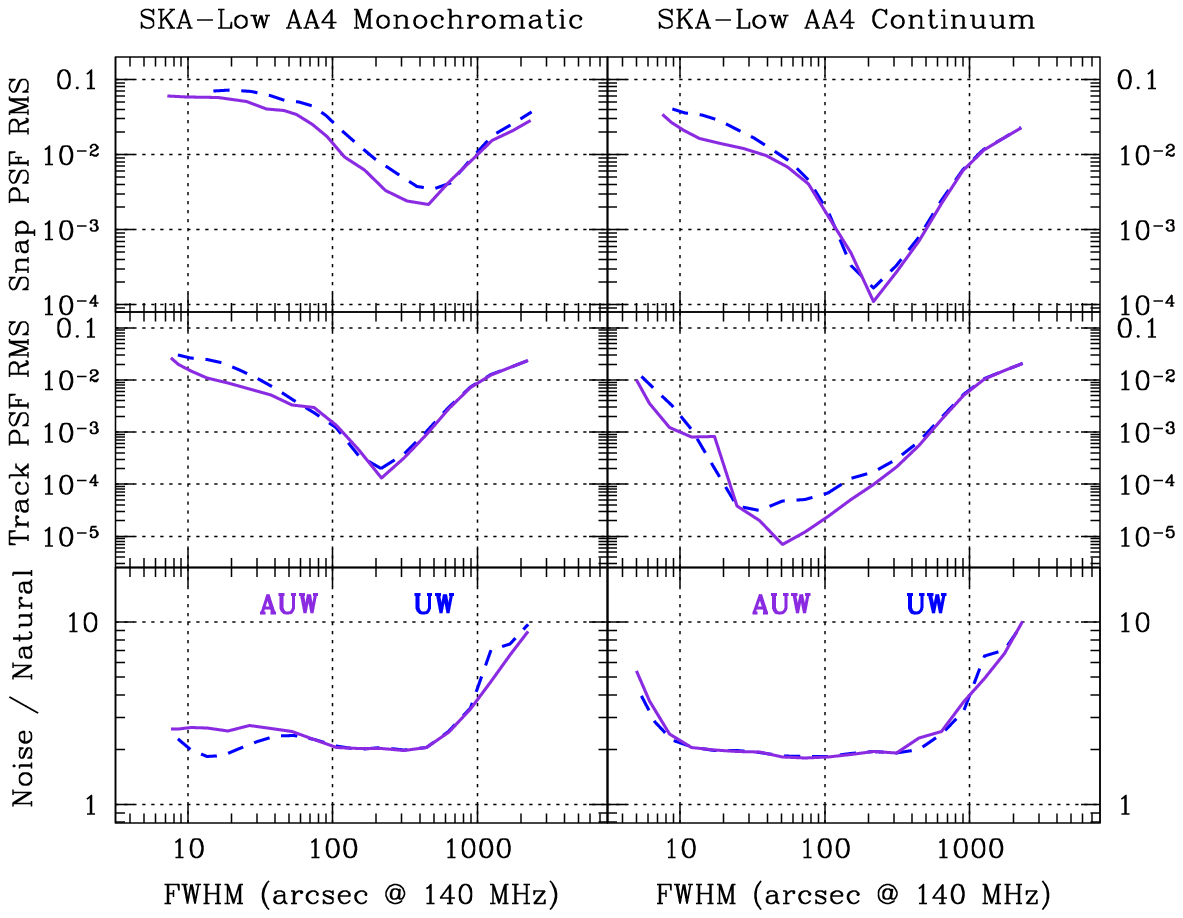}
      \caption{As in Figure \ref{fig:SKA-Mid} for SKA-Low.}
        \label{fig:SKA-Low}
   \end{figure*}
   
   \begin{figure*}
   \centering
    \includegraphics[bb=20 250 581 693,clip, width=14.5cm]{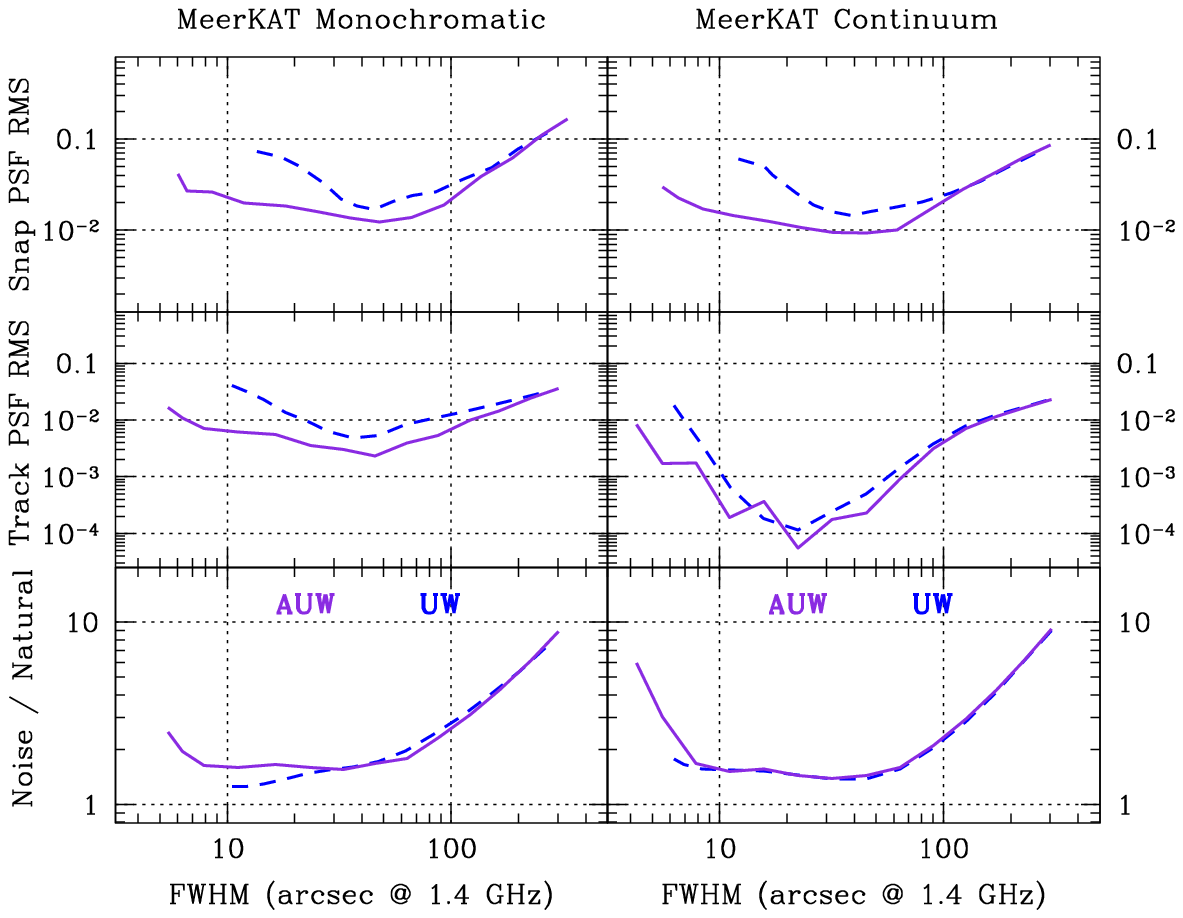}
      \caption{As in Figure \ref{fig:SKA-Mid}  for MeerKAT.}
        \label{fig:MeerKAT}
   \end{figure*}

   \begin{figure*}
   \centering
    \includegraphics[bb=20 250 581 693,clip, width=14.5cm]{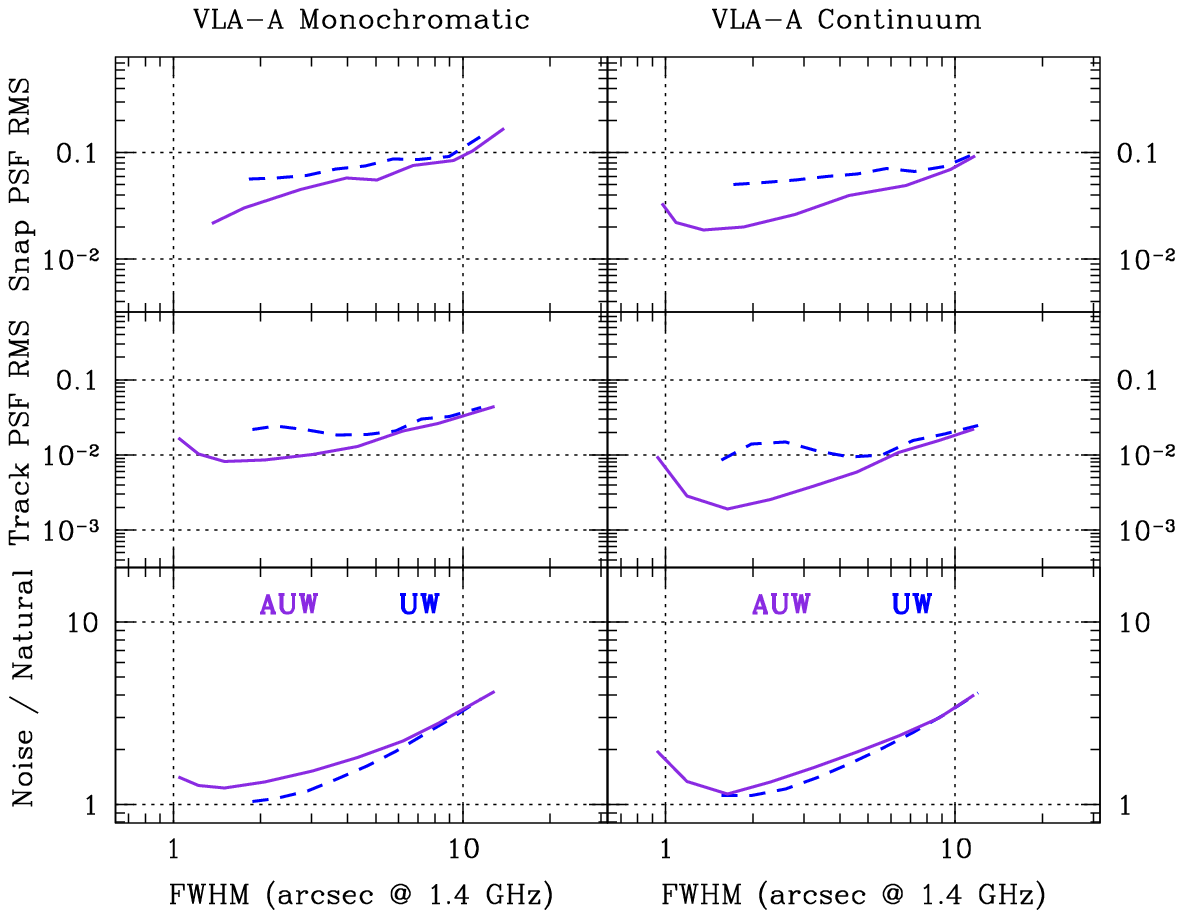}
      \caption{As in Figure \ref{fig:SKA-Mid}  for the VLA A-configuration.}
        \label{fig:VLA-A}
   \end{figure*}

  \begin{figure*}
   \centering
    \includegraphics[bb=20 250 581 693,clip, width=14.5cm]{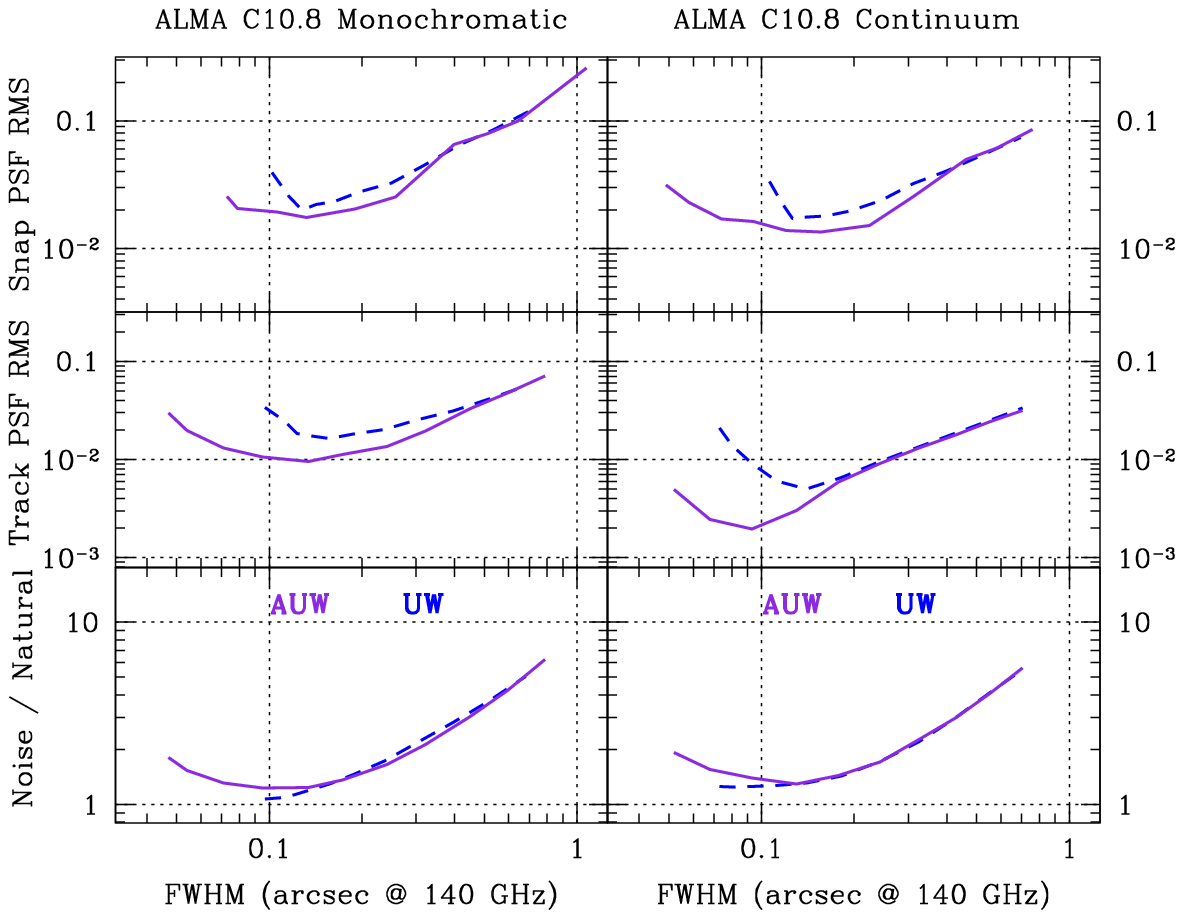}
      \caption{As in Figure \ref{fig:SKA-Mid}  for ALMA (Cycle 10-8 configuration).}
        \label{fig:ALMA}
   \end{figure*}

\end{appendix}
\end{document}